\documentclass[12pt]{article}
\usepackage[dvips]{graphicx}
\usepackage{amssymb}
\newcommand\be{\begin{equation}}
\newcommand\ee{\end{equation}}
\newcommand\fa{\begin{eqnarray}}
\newcommand\ffa{\end{eqnarray}}

\newcommand{\Z}{{\mathbb Z}}
\newcommand{\sgn}{\mathop{\rm sgn}}
\newcommand{\lP}{l_{\rm P}}
\begin{document}
{\renewcommand{\thefootnote}{\fnsymbol{footnote}}
\medskip
\begin{center}
{\LARGE  Isotropic Loop Quantum Cosmology with Matter II: The Lorentzian Constraint}\\
\vspace{1.5em} 
Franz Hinterleitner$^{a,b}$\footnote{ e-mail address: {\tt franz@physics.muni.cz}} and Seth Major$^{a,c}$\footnote{e-mail address: {\tt smajor@hamilton.edu}}\\  
\vspace{0.5em}  
$^a$Department of Physics, Hamilton College,\\  
Clinton, NY 13323, USA\\[2mm] $^b$Department of Theoretical  
Physics and Astrophysics,  Masaryk University,\\  
Kotl\'{a}\v{r}sk\'{a} 2, 611 37 Brno, Czech Republic\\[2mm]  
$^c$The Perimeter Institute\\ 
35 King Street North, Waterloo, Ontario, Canada N2J 2W9
\vspace{1.5em}  
\end{center}  }  
\setcounter{footnote}{0}  

\begin{abstract}  
The Lorentzian Hamiltonian constraint is solved for isotropic loop quantum cosmology coupled to a massless scalar field. As in the Euclidean case, the discreteness of quantum geometry removes the classical singularity from the quantum Friedmann models.  In spite of the absence of the classical singularity, a modified DeWitt initial condition is incompatible with a late-time smooth behavior.  Further, the smooth behavior is recovered only for positive {\em or} negatives times but not both. An important feature, which is shared with the Euclidean case, is a minimal initial energy of the order of the Planck energy required for the system to evolve dynamically. By forming wave packets of the matter field an explicit evolution in terms of an internal time is obtained.
\end{abstract}  

\section{Introduction}  
\label{sec:1}  

The recent paper \cite{MF} reports on a quantization of a Euclidean cosmology in which a  massless scalar field was coupled to a spatially flat  Friedmann-Robertson-Walker model.  Set in the framework of isotropic loop quantum gravity (ILQC) \cite{IL}, this model describes cosmological evolution in terms of a discrete time.  This discreteness is a direct  consequence of the kinematics of loop quantum gravity which predicts that geometric quantities such as area \cite{RSareavol,FLR,QGI}, volume \cite{RSareavol,L1,L2,QGII,RC}, and angle \cite{S} have a discrete spectrum.  In \cite{IL} the 
Hamiltonian constraint of quantum gravity, expressed in a discrete basis of volume eigenfunctions, and the matter Hamiltonian act on the state function of the coupled system.  This model avoids the classical singularity.  Due to the quantization of the inverse scale factor of the Friedmann model  
\cite{SF} the total Hamiltonian constraint vanishes at the classical singularity.  At small values of  the volume the wave function displays a distinct discrete behavior.   Nevertheless as the volume of the model grows, the wave function approaches a continuous function which is a solution of the asymptotic Wheeler - DeWitt differential  equation.  The wave function thereby meets the semi-classicality requirements of quantum  cosmology \cite{SC}.  

In the present paper we investigate the {\em Lorentzian} Hamiltonian constraint for the same massless scalar field model which is constructed along the prescriptions of the general theory \cite{QSD}, as it was introduced in \cite{IL}. In the spatially flat Friedmann model, however, the classical full Lorentzian constraint is proportional to the Euclidean one, so there is a quantum ambiguity and one can, as in recent work \cite{HL}, alternatively consider the Euclidean constraint operator proportional to the full one. In the semiclassical regime the results of both versions converge. The choice in this paper is closer to the full theory of quantum geometry \cite{QSD}.

We concentrate on three aspects of the ILQC framework.  First, the Hamiltonian constraint at early times leads to a consistency relation for initial data - the ``dynamical initial conditions" of \cite{Dyn}.  Although the Lorentzian constraint is higher order than the Euclidean constraint, we find the same relation in the model.    Second, to select an essentially unique solution one may require late time solutions to be smooth.  In such a ``pre-classical" \cite{Dyn} state the wave function at late times does not vary strongly on short intervals.  We find that this criteria again selects an essentially unique solution.  However, in contrast with the Euclidean model evolution backwards to negative time destroys the pre-classical condition.  Third,  an unexpected feature and, at the same time the main result of   \cite{MF}, is the occurrence of a threshold for dynamical evolution  of the model.  For the wave function to have a dynamical interpretation with respect to an internal time there is a minimal energy on the order of the Planck energy, concentrated initially in a volume of the size of a Planck volume.
 
 Scalar field quantum cosmology was first considered by Blyth and Isham \cite{BI}.  Canonically quantizing the reduced model, they explored the dynamics for several choices of time.  The current work is in sharp contrast to this older work in that the discreteness of ILQG manifests itself in early times and completely changes the status of initial conditions of the cosmological model.
 
 As we are dealing with  a generalization of the framework of \cite{MF} in section \ref{sec:2} we present the prerequisites very briefly .  In section \ref{sec:3} the discrete Hamiltonian constraint is solved numerically and the solutions are discussed.  In section \ref{sec:4} the asymptotic Wheeler - DeWitt equation and its solutions are constructed and compared with the Euclidean case. In section \ref{sec:5} an essentially unique wave function with a sufficiently smooth late time behavior to represent a classical universe, is singled out from the general solution of the Hamiltonian constraint. Section \ref{sec:6} contains a study of  possible wave packets in the scalar field and their  classical interpretation.
 
 \section{Prelimaries: Loop Quantum Cosmology}  
 \label{sec:2}  

We start with a quantization of the spatially flat Friedmann-Robertson-Walker model with a scalar field as matter source. The metric
\[  
 {\rm d}s^2=-{\rm d}t^2+a^2(t)\left[{\rm d}x^2+{\rm d}y^2+{\rm  d}z^2\right]  
\]  
represents a homogenous, isotropic model with one dynamical degree of freedom, represented, for example, by the scale factor $a(t)$ and its canonically conjugate momentum. In the framework of loop quantum gravity, applied to isotropic models \cite{IL}, the metric variable $a(t)$ is replaced by a triad variable $p$.  This triad variable can assume both signs according to the two possible orientations of a triad. By defining
\be
\label{p}
p:=a^2\sgn(a)
\ee
the domain of $a$ now extends to negative values. The conjugate momentum is
\be
\label{c}
c=\frac{1}{2}\dot a.
\ee 

Isotropic loop quantum cosmology \cite{IL} yields a discrete basis $|n\rangle$ of  volume eigenstates,   
\be
\label{Vol} 
\hat{V}|n\rangle=\left(\frac{1}{6}\gamma  \lP^2\right)^\frac{3}{2}\sqrt{(|n|-1)|n|(|n|+1)}\;|n\rangle =:  
V_{\frac{1}{2}(|n|-1)}|n\rangle,
\ee  
where the integers $n$ label discrete values of $p$ (or the scale factor)
\be \label{n} n=\frac{6}{\gamma l_P^2}p. \ee

In this basis the inverse scale factor operator, which is constructed independently from $a$, is also diagonal \cite{SC},   
\be \label{invsc} \widehat{|a|^{-1}}|n\rangle=  
16(\gamma\lP^2)^{-2}\left(\sqrt{V_{\frac{1}{2}|n|}}-  
\sqrt{V_{\frac{1}{2}|n|-1}}\right)^2|n\rangle. \ee   
This, in contrast to the inverse of the scale factor operator, is a   
densely defined operator on the Hilbert space spanned by the basis   
$\{|n\rangle\}$.  

For the purposes of evaluating the Hamiltonian constraint the most important feature of this construction is that the eigenvalue zero of the volume is threefold degenerate, as can be seen from (\ref{Vol}), it vanishes in $|0\rangle$ and in  $|\pm1\rangle$, whereas the eigenvalue of $\widehat{|a|^{-1}}$ is zero only on $|0\rangle$. The latter state assumes the role analogous to  the classical singularity $a=0$. The vanishing of the inverse scale factor is a pure quantum feature, in sharp contrast to the divergence of the classical $a^{-1}$ for $a=0$. 

An arbitrary state $|s\rangle$ of the cosmological model can be expressed as \be |s\rangle=\sum_{-\infty}^\infty s_n|n\rangle.  \ee In the ILQC framework the full Lorentzian Hamiltonian constraint $\hat{H}|s\rangle=0$ assumes the form of a difference equation for the coefficients $s_n$ \cite{IL}  
\begin{eqnarray} \label{Hgrav}   
&&3(\gamma\kappa   
l_P^2)^{-1}\left[\mbox{$\frac{1}{4}$}(1+\gamma^{-2})\sgn(n+8)  
\left(V_{\frac{1}{2}|n+8|}-V_{\frac{1}{2}|n+8|-1}\right)  
k_{n+8}^+k_{n+4}^+s_{n+8}  
\right.\nonumber\\  
&&-\sgn(n+4)\left(V_{\frac{1}{2}|n+4|}-  
V_{\frac{1}{2}|n+4|-1}\right)s_{n+4}\nonumber\\  
&&-2\sgn(n)\left(V_{\frac{1}{2}|n|}-V_{\frac{1}{2}|n|-1}\right)  
(\mbox{$\frac{1}{8}$}(1+  
\gamma^{-2})(k_n^-k_{n+4}^++k_n^+k_{n-4}^-)-1)s_n\nonumber\\  
&&-\sgn(n-4)\left(V_{\frac{1}{2}|n-4|}-V_{\frac{1}{2}|n-4|-1}\right)s_{n-4}\\  
&&\left.+\mbox{$\frac{1}{4}$}(1+\gamma^{-2})\sgn(n-8)  
\left(V_{\frac{1}{2}|n-8|}-V_{\frac{1}{2}  
|n-8|-1}\right)k_{n-8}^-k_{n-4}^-s_{n-8}\right]=0.\nonumber  
\end{eqnarray}  
Thanks to the sign factors the coefficient $s_0$ of the singular state drops out and can be set equal to zero. In this sense the model based on loop quantum gravity is singularity-free. 

The $k$'s, coming from the extrinsic curvature contribution to the ``kinetic" term of the Hamiltonian, have the following expressions in terms of volume eigenvalues  
\begin{eqnarray}   
    \label{k+}   
    k_n^+=&&\!\!\!\!\!\!3(\gamma  
    l_P^2)^{-3}\left[\left(V_{\frac{1}{2}(|n+1|-1)}-  
    V_{\frac{1}{2}(|n-3|-1)}\right)  
    \left(V_{\frac{1}{2}|n-3|}-V_{\frac{1}{2}|n-3|-1}+  
    V_{\frac{1}{2}|n+1|}-V_{\frac{1}{2}|n+1|-1}\right)  
    \right.\nonumber\\  
    &&\!\!\!\!\!\!-\left.\left(V_{\frac{1}{2}(|n-1|-1)}-  
    V_{\frac{1}{2}(|n-5|-1)}\right)\left(  
    V_{\frac{1}{2}|n-5|}-V_{\frac{1}{2}|n-5|-1}+  
    V_{\frac{1}{2}|n-1|}-V_{\frac{1}{2}|n-1|-1}\right)\right],  
\end{eqnarray}  
and  
\begin{eqnarray}   
    \label{k-}   
    k_n^-=&&\!\!\!\!\!\!3(\gamma  
    l_P^2)^{-3}\left[\left(V_{\frac{1}{2}(|n+5|-1)}-  
    V_{\frac{1}{2}(|n+1|-1)}\right)\left(V_{\frac{1}{2}|n+1|}-  
    V_{\frac{1}{2}|n+1|-1}+V_{\frac{1}{2}|n+5|}-V_{\frac{1}{2}|n+5|-1}\right)  
    \right.\nonumber\\  
    &&\!\!\!\!\!\!-\left.\left(V_{\frac{1}{2}(|n+3|-1)}-  
    V_{\frac{1}{2}(|n-1|-1)}\right)\left(  
    V_{\frac{1}{2}|n-1|}-V_{\frac{1}{2}|n-1|-1}+  
    V_{\frac{1}{2}|n+3|}-V_{\frac{1}{2}  
    |n+3|-1}\right)\right]  
\end{eqnarray}  
and are subject to the identities \be \label{id1} k_n^+=k_{-n}^- \ee  
and \be \label {id2} k_{n+4}^+=k_n^-.  \ee These identities make only  
one kind of $k$'s necessary; we choose $k^+$.  
  
The model under consideration includes a massless scalar field  
$\phi$.  Its classically Hamiltonian in the Friedmann-Robertson-Walker metric is $H_\phi = \frac{1}{2} p_\phi^2 a^{-3}$.  When quantized it becomes \cite{MF} 
\be  
\label{Hmatter} \hat{H}_\phi(n)=-\frac{1}{2}\hbar^216^3  
(\gamma\lP^2)^{-6} \left(\sqrt{V_{\frac{1}{2}|n|}}-  
\sqrt{V_{\frac{1}{2}|n|-1}}\right)^6\; \frac{{\rm d}^2}{{\rm  
d}\phi^2},
\ee
which is the energy operator per unit coordinate volume. Here the quantization of $|a|^{-1}$ (\ref{invsc}) is crucial for it renders the classical Hamiltonian finite when $a=0$. In the total Hamiltonian $\hat{H}_\phi$ couples with a sign factor $\sgn(n)$ to the gravitational part.  
  
As in \cite{MF} we assume the states of the coupled system are of  
the form \be \label{s} |s\rangle=\sum_{-\infty}^\infty  
s_n(\phi)\,|n\rangle, \ee where the dependence on $\phi$ is contained  
in the coefficients $s_n$ of the quantum geometry basis vectors  
$|n\rangle$.  In our calculations we assume that the $\phi$-dependence  
of the state vector is given by an eigenfunction $\chi$ of the matter  
Hamiltonian, characterized by $\omega$, \be \label{chi}  
s_{n}(\phi)=:\check{s}_n\chi_\omega(\phi):=  
\check{s}_n\,e^{i\frac{\omega}{\hbar}\phi}.  \ee As we see in the next  
section, this ansatz results in a finite difference  
equation for the model.  
  
\section{The Hamiltonian constraint equation}  
\label{sec:3}   
  
The difference equation resulting from the Hamiltonian constraint (\ref{Hgrav}) is of order 16 so that the naive expectation is that the solution should contain 16 free parameters.  The solution decomposes into four independent series of coefficients, namely $\check{s}_{4m+i}$, $i=0,\ldots,3$, $m\in\Z$, with four free parameters each. In the series with $i=0$ $\check{s}_0$ drops out, because its coefficient in the gravitational hamiltonian contains sgn(0) and $\hat{H}_\phi(0)$ is the zero operator by virtue of its construction in terms of $\widehat{|a|^{-1}}$ \cite{IL,AS}. So, for example, for $n=8$, the Hamiltonian constraint (\ref{Hgrav}) with the matter Hamiltonian, does not relate $s_0$ to $s_4,\ldots,s_{16}$, but instead gives a {\em consistency condition} for $s_4,\ldots,s_{16}$, reducing the number of free parameters from four to three. This $i=0$ series is considered as fundamental.  By applying the ``pre-classicality condition" at late times we are able to pick out essentially unique solutions to the other series.  The idea is that, with the fundamental series, the other three series are selected by smooth interpolation.    Thus, the 16 free parameters are reduced to 3.

To facilitate the handling of the difference equation we simplify it by introducing rescaled coefficients   
\be   
\label{t}   
t_n:=\left(V_{\frac{1}{2}|n|}-V_{\frac{1}{2}|n|-1}\right)\check{s}_n  
\ee   
and use the abbreviation  
\be  
\beta:=\mbox{$\frac{1}{4}$}(1+\gamma^{-2}).  
\ee  
To avoid confusion, we remind the reader that $s_n$ denote the full coefficients, $\check{s}_n$ the $\phi$-independent parts and $t_n$ the rescaled $\check{s}_n$ used for solving the Hamiltonian constraint
\be \label{H} (\hat{H}+\sgn(n)\hat{H}_\phi)\sum s_n(\phi)|s\rangle=0. \ee
The sign factor in front of the matter Hamiltonian as well as the signs in (\ref{Hgrav}) give rise to a relative sign between the two (classically disjoint) sectors with $p>0$ and $p<0$ (see \cite{IL}). They render the above constraint equation time symmetric, provided $\hat{H}_\phi(n)=\hat{H}_\phi(-n)$.  For the moment,  we consider only positive values of $n$.  

\begin{figure}[ht]  
\begin{center}  
\includegraphics[height=8.5cm]{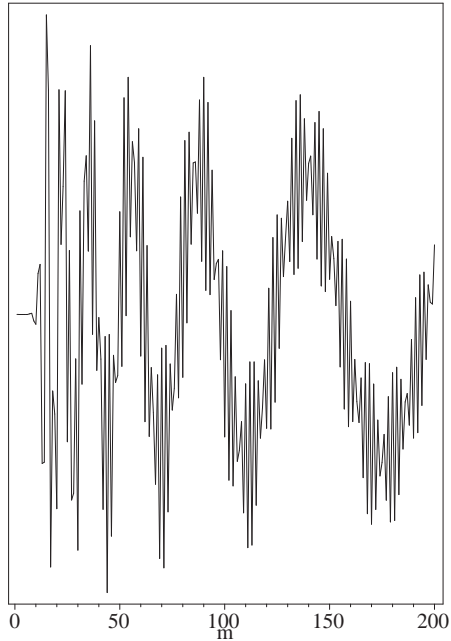}  
\end{center}  
\caption{ \label{a200} A solution $\check{s}_m$ of  
(\ref{de2}) with initial values $t_4=1$, $t_8=t_{12}=0$, $\gamma=0.13$, and $m=4n$ so that $0\leq n\leq 800$. The points $\check{s}_m$ of the wave function are connected by lines to clearly display the oscillatory character.}   
\end{figure}

By inserting $n=8$ into the total Hamiltonian constraint equation $(\hat{H}+\hat{H}_\phi)\sum s_n(\phi)|s\rangle=0$ we obtain $t_{16}$ in terms  of $t_4$, $t_8$ and $t_{12}$,  

\begin{eqnarray}   
    \label{de1}   
    t_{16}=&&\!\!\!\!\!\!\frac{1}{\beta   
    k_{12}^+k_{16}^+}\left\{t_{12}-\left[2-\rule{0mm}{8mm}  
    \beta\left[(k_{12}^+)^2+(k_8^+)^2\right]+\right.\right.\\  
    &&\!\!\!\!\!\!\left.\left.  
    \frac{2048  \omega^2}{3 \hbar \gamma^{5} l_P^{8}}  
    \frac{\left(\sqrt{V_4}-\sqrt{V_3}\right)^6}  
    {V_4-V_3}\right]t_8+t_4  
    \right\}.\nonumber   
\end{eqnarray}  
(This is the above-mentioned consistency condition.) The general form of the difference equation for $n\geq12$ is  
\begin{eqnarray}   
    \label{de2}   
    t_{n+8}=&&\!\!\!\!\!\!\frac{1}{\beta   
    k_{n+8}^+k_{n+4}^+}\left\{t_{n+4}-\left[2-\rule{0mm}{10mm}  
    \beta\left[(k_{n+4}^+)^2+(k_n^+)^2\right]+\right.\right.\\  
    &&\!\!\!\!\!\!\left.\left.  
    \frac{2048 \omega^2}{3 \hbar \gamma^{5} l_P^{8}}  
    \frac{\left(\sqrt{V_{\frac{n}{2}}}-\sqrt{V_{\frac{n}{2}-1}}\right)^6}  
    {V_{\frac{n}{2}}-V_{\frac{n}{2}-1}}\right]t_n+t_{n-4}-  
    \beta k_{n-4}^+k_n^+\cdot t_{n-8}\right\}.\nonumber   
\end{eqnarray}  
Equation (\ref{de2}) is complicated enough (the worst complications are hidden in the $k$'s) so that numerical methods are useful.  In figure \ref{a200} we plot the numerical solution for the initial  
conditions $t_4=1$, $t_8=t_{12}=0$ and a parameter $\omega$ such that $\frac{2048}{3}\frac{\omega^2}{\hbar\gamma^5 l_P^{10}}=10^7$. In this numerical solution we use the value of $\gamma = \frac{\ln 2}{\pi \sqrt{3}}\sim 0.13$ for the Immirzi parameter.\footnote{This value merges from isolated horizons work of \cite{ABK} and in a general model of gravitational statistical mechanics \cite{MS}.}  Although the solutions were obtained from the initial values $t_4=1$, $t_8=t_{12}=0$ the solution is generic in that the solution is qualitatively the same when the initial values are varied by as much as $10^4$.

In contrast to the Euclidean case, persistent short-wavelength oscillations continue to late times -- even deep into what we would expect to be the semiclassical regime. These short-wavelength oscillations appear to be superimposed on a smooth function of the same shape as the Euclidean wave function.   

\section{The continuum limit}  
\label{sec:4}  
  
The physical implications of the Lorentzian Hamiltonian constraint can be deduced without a detailed solution of (\ref{de2}), but from the asymptotic smooth mean-value functions, which are approximated for  large values of $n$ (or the volume).  These functions are solutions of a Wheeler-DeWitt differential equation, which can be derived from (\ref{de2}) as continuum limit.  To distinguish between discrete and continuous quantities we return the continuous variable $p$, whose relation to $n$ is given by (\ref{p}). 

\begin{figure}[ht]  
\begin{center}  
 \includegraphics[width=8cm,angle=270]{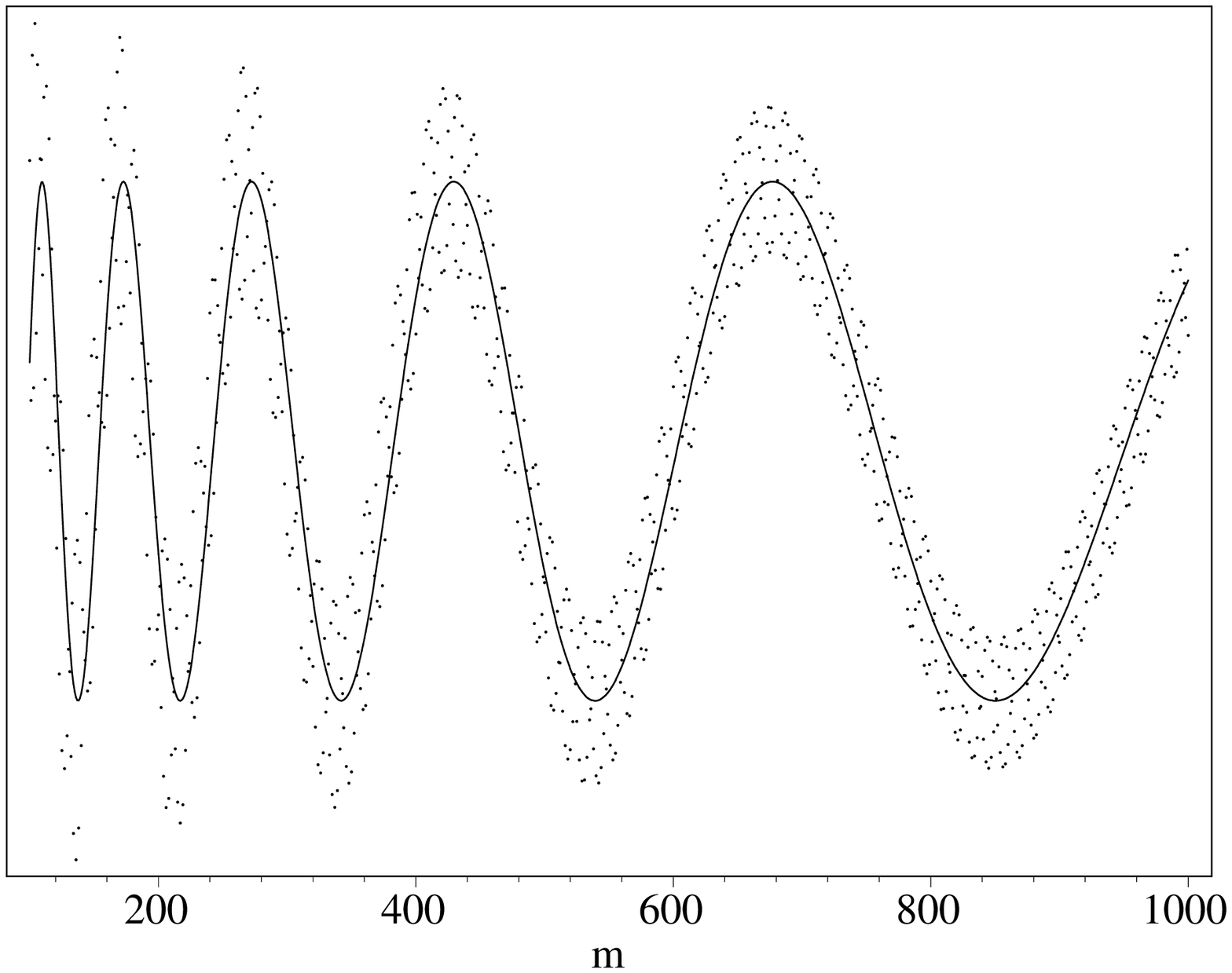}  
\end{center}  
\caption{\label{match} The numerical result (points) and the asymptotic solution (solid line) for the $\phi$-independent part   $\check{s}_{4m}$ of the wave function. The conditions are the same  as in figure \ref{a200}  but here  $100\leq m \leq 1000$.} On this scale the high frequency oscillations are visible in the ``cloud" of points around the asymptotic solution.
\end{figure}

We also need some  asymptotic expansions:   
\begin{eqnarray}  
k_{n}^+\sim1+\frac{3}{8}\frac{1}{n^2}=1+\frac{(\gamma  
l_P^2)^2}{96}\frac{1}{p^2},\\  
V_{\frac{1}{2}n}-V_{\frac{1}{2}n-1}\sim 24^{-\frac{1}{2}}(\gamma  
l_P^2)^{\frac{3}{2}}n^{\frac{1}{2}}=\mbox{$\frac{1}{2}$}\gamma  
l_P^2\sqrt{p}, \\  
\left(\sqrt{V_{\frac{1}{2}n}}-\sqrt{V_{\frac{1}{2}n-1}}\right)^6\sim  
\left(\mbox{$\frac{3}{128}$}\right)^{\frac{3}{2}}(\gamma  
l_P^2)^{\frac{9}{2}}n^{-\frac{3}{2}}=\left(\mbox{$\frac{1}{4}$}\gamma  
l_P^2\right)^6p^{-\frac{3}{2}}.  
\end{eqnarray}  
Assuming the $\phi$-dependence to be given by the function   
$\chi_\omega(\phi)$ of equation (\ref{chi}), we rewrite the Hamiltonian   
constraint (\ref{de2}) in the form  
\begin{eqnarray}  
&&\beta\left(k_{n+8}^+k_{n+4}^+\,   
t_{n+8}-\left[\left(k_{n+4}^+\right)^2+  
\left(k_n^+\right)^2\right]t_n+k_n^+k_{n-4}^+\,   
t_{n-8}\right)\\  
&&-\left(t_{n+4}-2t_n+t_{n-4}\right)  
=-\frac{2048\omega^2}{3 \hbar \gamma^{5}  l_P^8}  
\frac{\left(\sqrt{V_{\frac{1}{2}n}}-\sqrt{V_{\frac{1}{2}n-1}}\right)^6}  
{V_{\frac{1}{2}n}-V_{\frac{1}{2}n-1}}\:t_n,\nonumber  
\end{eqnarray}  
so that we have two kinds of differences on the left-hand side, differences of products of $t$'s by $k$'s and differences of $t$'s alone.  In the continuous limit for large $n$ the later ones are second  
derivatives with respect to $p$. Thanks to the fact that the $k$'s are, to order $1/n^{2}$,  
equal to one, the former differences are also approximated  by the second derivative in leading order.  Hence the continuous limit of the left-hand side becomes 
\[ (64\beta-16)\frac{{\rm d}^2t}{{\rm  
d}n^2}=\frac{4l_P^4}{9}\frac{{\rm d}^2t(p)}{{\rm d}p^2}.  \]   
in which we write the asymptotic form of $t_n$ as $t(p)$. (For the moment we pretend that $n$ is continuous.) Together with the leading term of the right-hand side this gives a Cauchy-Euler  
equation of order 2
\be
\label{diff} 
p^{2} \, \frac{{\rm d}^2t(p)}{{\rm  
d}p^2}+\frac{3\kappa\omega^2}{4l_P^4}\:t(p) =0
\ee with the solutions  
\be \label{tsolutions} t(p)=p^{\frac{1}{2}}e^{\pm i\Omega\ln p}, \ee 
where $\Omega$ is  
given by 
\be  
\Omega=\mbox{$\frac{1}{2}$}\sqrt{3\kappa\omega^2l_P^{-4}-1}.  
\ee
From  this solution we may construct the continuous limit $s_\omega(p,\phi)$  of $s_{n}(\phi)$ from the relation $s_{n}(\phi)=t_n/(V_{n/2}-V_{n/2-1})\,\chi_\omega(\phi)$.  Taking into account that $V_{\frac{1}{2}n}-V_{\frac{1}{2}n-1}$ goes as $n^{\frac{1}{2}}$ in leading order we have 
\be 
\label{asymp} 
s_\omega(p,\phi)=e^{\pm i\Omega\ln p}e^{\pm  i \frac{\omega}{\hbar}\phi}.  
\ee
The continuous limit of the wave  function is thus the same as in the Euclidean case \cite{MF}.  The same is true for the existence of the critical value $\omega_{\rm  crit}=l_P^2/\sqrt{3\kappa}$.  The behavior of the wave function is crucially different according to the value of $\omega$.  The wave function only displays the asymptotic oscillatory behavior when $\omega>\omega_{\rm crit}$.  The value $\omega_{\rm  crit}$ corresponds to a threshold value of an initial energy of the  order of the Planck energy, concentrated in a volume of the order $l_P^3$ at the classical singularity (see \cite{MF}).\footnote{The precise value may be modified by quantum ambiguities \cite{QA,In}.} Only if such an amount of energy or more is present, dynamical evolution of  the model is possible. Figure \ref{match} shows the approximation of an asymptotic wave function of the type (\ref{asymp}) by a solution of the Hamiltonian  constraint (\ref{de2}) for $\phi=0$.

\section{Pre-classicality}  
\label{sec:5}  
\subsection{Finding the pre-classical solution}  

In \cite{Dyn} it is argued that in loop quantum cosmology there always exists a ``pre-classical" solution which is smooth at late times. To find this solution among the three-parameter family of the general solution, it is convenient to look at a different asymptotic limit than achieved in equation (\ref{diff}).   

This limit may be found in \cite{Dyn}.  But to keep our discussion self-contained we outline it here. The approximation is based on replacing the $k$'s by their asymptotic value of 1 and by considering the matter contribution to the constraint equation as small and constant. Indeed, $(\sqrt{V_{n/2}}-\sqrt{V_{n/2-1}})^6(V_{n/2}-V_{n/2-1})^{-1}$ goes as $n^{-2}$ for $n\gg1$, so this approximation is valid on finite ranges of $n$, determined by the condition that $n^{-2}$ does not change much on such a range.  We define  
\be 2P:=\frac{2048\omega^2}{3\hbar\gamma^5l_P^8}\frac{(\sqrt{V_{\frac{1}{2}n}}-\sqrt{V_{\frac{1}{2}n-1}})^6}{V_{\frac{1}{2}n}-V_{\frac{1}{2}n-1}},\ee  
which is much smaller than 1 for large $n$ and use the index $m=n/4$ in the following.   

Considering $P$ as constant we obtain an asymptotic difference equation with constant coefficients instead of a differential equation,  
\be \beta t_{m+2}-t_{m+1}+2(1-\beta+P)t_m-t_{m-1}+\beta t_{m-2}=0, 
\ee
conserving discreteness.  An exponential ansatz $t_m=e^{im\vartheta}$ leads to a quadratic equation for $\cos\vartheta$ with the approximate solutions  
\be \cos\vartheta_0\approx1+\frac{P}{1-4\beta}\hspace{1cm}\mbox{and}\hspace{1cm}\cos\vartheta_1\approx\frac{1-2\beta}{2\beta}-\frac{P}{1-4\beta}. \ee
From this we obtain 4 solutions, $e^{\pm im\vartheta_0}$ and $e^{\pm im\vartheta_1}$, where $\vartheta_0$ goes asymptotically to zero and $\vartheta_1$ does not. In consequence, $e^{\pm im\vartheta_0}$ approximate a constant in a range of $m$ which is much smaller than the wavelength of the main oscillation.  On an arbitrary long range these solutions approximate a smooth solution of (\ref{diff}). This solution is called pre-classical.  This analysis shows that the asymptotic behavior of the three linearly independent functions obtained from the three initial parameters is of the form  
\be f_{\rm WDW}^{(i)}+a^{(i)}\cos\vartheta_1m+b^{(i)}\sin\vartheta_1m,\hspace{2cm}i=1,2,3,\ee  
where $f_{\rm WDW}^{(i)}$ is a real, $p$ (respectively $n$)-dependent part of a solution of the asymptotic Wheeler-DeWitt equation. For any linearly independent triple of such functions there is obviously a unique, nontrivial linear combination in which the sine and cosine of $\vartheta_1m$ cancel.

In the numerical solution extreme fine tuning of the initial values is necessary to find smooth solutions.\footnote{Changing the initial values by one part in $10^{11}$ will re-introduce the oscillatory ``quantum foam" at later times.}  It is better to determine these values indirectly. With our set of 3 initial values we can produce a variety of solutions, whose average curves, as shown in figure 2, are in arbitrary phase relations with one another. These smooth functions form the 2-dimensional space of solutions of the asymptotic differential equation (\ref{diff}). 

To determine the pre-classical solution we evolve smooth initial data for sufficiently large values of $n$ backwards to $n=0$. For this purpose we may choose any set of 4 smooth data; every set  gives rise to a solution of (\ref{diff}) in the asymptotic region. Back evolution yields fairly smooth functions back to a early domain in $n$. In this domain and earlier the functions begin to oscillate rapidly and, in general, they do not fulfill the consistency condition (\ref{de1}) for $t_4$ to $t_{16}$ and thus for $\check{s}_0$ to $\check{s}_{16}$. 

Constructing two linearly independent asymptotic solutions, we obtain a basis of solutions of the asymptotic Wheeler-DeWitt differential equation, from which it is possible to construct that linear combination, which fulfills the consistency condition. This combination yields the initial data $\check{s}_4$, $\check{s}_8$, and $\check{s}_{12}$, whose forward evolution, shown in figure 3, is the pre-classical solution for the given values of $\gamma$ and the field energy.

\begin{figure}[ht]  
\begin{center}  
\includegraphics[height=10cm,angle=270]{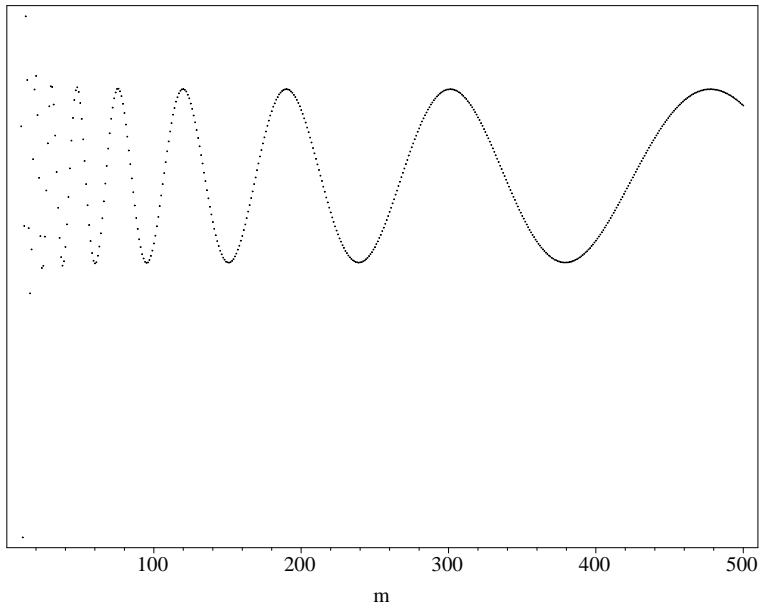}  
\end{center}  
\caption{\label{abb3} The pre-classical solution for the same values of $\gamma$ and the field energy as in figure 1 for $n$ from 40 to 2000. For smaller values of $n$ the amplitude increases, the largest coefficient $\check{s}_4$ is equal to 150222.0.}  
\end{figure}  

As a practical matter the essential uniqueness of the pre-classical solution has to be understood within a certain latitude. Late time solutions are very sensitive to small changes in the initial data for $n=4,8,12$. For instance, the solution depicted in figure 3 is not the optimal one, although it appears sufficiently smooth in the figure. The solution arises as a linear combination of two pre-classical functions, evolved back from the late time initial values $(\check{s}_{1988},\check{s}_{1992},\check{s}_{1996},\check{s}_{2000})$ equal to $(1,1,1,1)$ and $(0.7,0.8,0.9,1)$, respectively. In this domain of $n$ these linearly combined initial data are a sufficiently good approximation to pre-classical wave functions (\ref{tsolutions}), so that the resulting graphic looks perfectly smooth. By a slight modification, they could be adjusted to a smooth asymptotic solution of the form (\ref{tsolutions}) in order to improve the pre-classicality of the resulting function.   

One might hope that a symmetry principle  might select the pre-classical wave function.  For instance we might consider data which is symmetric or antisymmetric about the time of the classical singularity.  As we see in the next section though, neither of these cases select a pre-classical solution.  

\subsection{Evolution through the classical singularity}  

The absence of singularities in loop quantum cosmology enables the wave function to evolve through the time of the classical singularity into a domain with negative values of $n$. Evolving the pre-classical wave function backwards reveals a further essential difference between the Euclidean and Lorentzian cases: When $n$ approaches zero from the positive side, the amplitude of the wave function begins to increase. Beyond zero it continues growing and after large and rapid oscillations the wave function settles down to the superposition of the two oscillations, already familiar from figures \ref{a200} and \ref{match}.  Note, however, that for negative $n$'s the wavefunction has a very large amplitude compared to positive $n$'s. For our values of $\gamma$ and the field energy, the ratio between the amplitudes for negative and positive values of $n$ is of the order of $10^{18}$.  Naively, the wave function for negative and positive $n$ suggests that there is vanishing probability of the pre-classical cosmology at positive $n$.
  
\begin{figure}[ht]  
\begin{center}  
\includegraphics[height=10cm,angle=270]{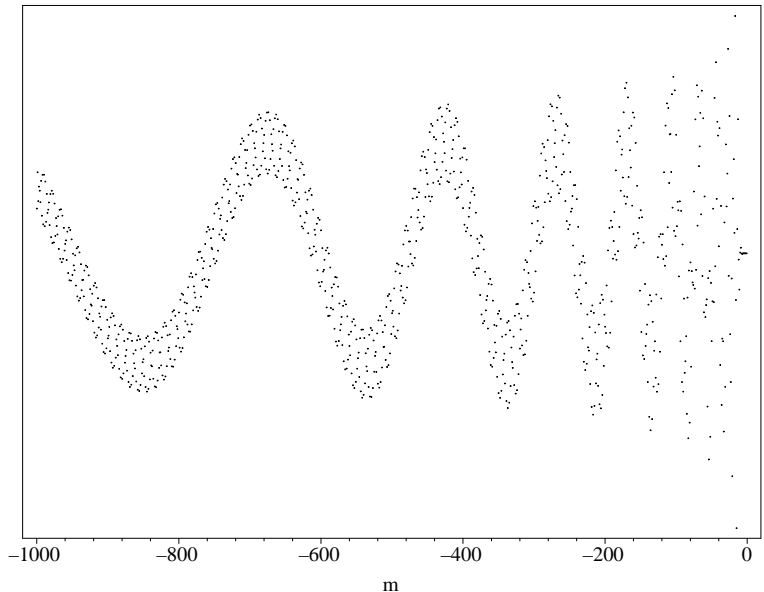}  
\end{center}  
\caption{\label{abb4} Continuation of the above pre-classical wave function to $-1000\leq m\leq-1$.}  
\end{figure}  

Near the classical singularity, Lorentzian wave functions display large oscillations. The larger the energy of the scalar field, the longer is the ``quantum regime" of these oscillations at positive $n$'s and the later the pre-classical behavior sets in. 

\section{Wave packets and dynamics}  
\label{sec:6}
  
So far, the functional dependence of the wave function on the field  
was assumed to be of the form $\exp(\pm i\frac{\omega}{\hbar}\phi)$.   
The field being massless and spatially constant, $\phi$ is formally  
equivalent to a configuration variable of a particle with gravity  
acting as a one-dimensional potential.  In this analogy the critical  
energy separates free states from bound ones (although we did not  
define a measure on $n$ to give the wave functions a proper meaning of  
a probability amplitude at a certain value of $\phi$).  
  
To investigate wave packets in $\phi$, made from superpositions of  
$\chi_\omega$'s for different $\omega$'s,  we use  
\be   
\label{pack}  
s_n(\phi)=\check{s}_n\cdot\int_{-\infty}^\infty\!\!{\rm  
d}\omega\,e^{-\lambda(\omega-\omega_0)^2}e^{-i\frac{\omega}{\hbar}\phi}  
\ee  
with $\omega_0\gg\omega_{\rm crit}$ and  
$\lambda\gg(\omega_0-\omega_{\rm crit})^{-2}$.  The two conditions  
assure that the contribution from $\omega\leq\omega_{\rm crit}$ is  
negligible and no under critical wave functions are included.  
  
The Wheeler - DeWitt equation being linear, we obtain in the   
continuous limit a superposition of functions (\ref{asymp})  
\begin{eqnarray*}   
s(p,\phi)&=&(2\pi\lambda\hbar^2)^{-\frac{1}{4}}  
\int_{-\infty}^\infty{\rm d}\omega\,e^{-\lambda(\omega-\omega_0)^2}  
\,e^{-i\frac{\omega}{\hbar}\phi}  
\,e^{\frac{i}{2}\sqrt{\left(\frac{\omega}{\omega_{\rm crit}}\right)^2-1}\,  
\ln p}\\  
&&\approx\int_{-\infty}^\infty{\rm d}\omega\,  
e^{-\lambda(\omega-\omega_0)^2}\,  
e^{-\frac{i\omega}{\hbar}\phi}\,e^{\frac{i}{2}\frac{\omega}{\omega_{\rm   
crit}}\ln p}.  
\end{eqnarray*}  
With normalization in $\phi$ with respect  to the natural inner product  
$$ \langle s|t\rangle=\int_{-\infty}^\infty{\rm   
d}\phi\,s^*(p,\phi)t(p,\phi) $$  
this leads to the following modulated Gau\ss\ function in $\phi$  
\be \label{pack2} s(p,\phi) \approx (2\pi\lambda\hbar^2)^{-\frac{1}{4}}  
 e^{-\frac{1}{4\lambda}(\frac{\ln p}{2\omega_{\rm   
crit}}-\frac{\phi}{\hbar})^2}e^{i\omega_0(\frac{\ln p}{2\omega_{\rm   
crit}}-\frac{\phi}{\hbar})} \ee  
with a maximum at $\phi=\frac{\hbar\ln p}{2\omega_{\rm crit}}$ and a   
width $2\sqrt{\lambda}$.   
  
Considering now $p$ as time variable, we can calculate $p$-dependent   
expectation values of $\phi$ and the (kinetic) field energy.  For   
$\phi$   
\be   
\label{phi}   
\langle\phi(p)\rangle=\frac{\hbar}{2\omega_{\rm crit}}\ln p,  
\ee  
showing a growth during the expansion of the universe.  This is an  
indication that in a more sophisticated model with an inherent notion  
of particles, like a massive or a spatially non-constant field, one  
could expect particle creation.  
The energy expectation value  
\be  
\label{en}  
E_\phi(p):=\langle\hat{H}_\phi(p)\rangle=  
\langle-\frac{\hbar^2}{2}\,p^{-\frac{3}{2}}\,\frac{{\rm d}^2}{{\rm  
d}\phi^2}\rangle=\mbox{$\frac{1}{2}$}\left(\omega_0^2+1/4\lambda\right)\,  
p^{-\frac{3}{2}},    
\ee  
shows a decreasing field energy per unit  
coordinate volume.  Remembering the relation (\ref{p}) between the  
discrete counterpart $n$ of $p$ and the scale factor $a$ we find that  
the energy scales as $V^{-1}$.  The energy density per physical  
volume goes therefore as $V^{-2}$.  
  
In the above interpretation, when the radius of the universe is   
considered as internal time, there is no problem with superpositions   
of expanding and contraction universes or universes going backwards   
in time, as it would be, when $\phi$ acted as time. So the choice of   
$n$ or $p$ seems to be much more natural and less problematic -- the wave functions and expectation values of the matter field evolve with the scale factor.   

\section{Conclusions}  
\label{sec:7}  

Comparing the present Lorentzian with the Euclidean Hamiltonian constraint, one notices three similarities: (1) The asymptotic continuous wave function is of the same form. (2)  There is the same minimal initial energy (up to quantum ambiguities in both cases \cite{QA}) for dynamical evolution of the wave function.  (3) The extension of the quantum regime between the classical singularity and the beginning of a smooth, pre-classical evolution increases with growing matter energy.  Therefore the semiclassical limits of the wave functions coincide. This fact confirms that the difference between the two versions of the constraint is a quantum ambiguity of the spatially flat Friedmann model.     

In the quantum regime,  the models differ essentially from each other. The Euclidean wave function is small near the singularity, thus satisfying what one could call a ``modified DeWitt initial condition". The original DeWitt condition ensures that the wave function in standard quantum cosmology stays away from the singularity. The Lorentzian wave function, on the other hand, although being equal to zero at the singular state, is large in the immediate vicinity of the classical singularity.

In both cases it is possible to extend the principle series beyond the singularity to negative values of the discrete internal time $n$. In the Euclidean case there is only one solution (up to a normalizing factor), which is smooth and symmetric with respect to $n$. In the Lorentzian case the unique\footnote{Up to a small uncertainty, see the remark at the end of section 5.1.  This is in addition to the normalizing factor.} smooth solution is embedded in a three-parameter family of Planck-scale varying solutions. Furthermore, pre-classicality is only one-sided, i.e. the wave function is pre-classical, at best, for $n>0$ or $n<0$. 

Thus, if we assume that evolution from arbitrary negative to arbitrary positive values of the internal time is possible, we come to the following scenario. Whenever a pre-classical universe is to emerge from the considered model, it is preceded by a non-pre-classical ``fuzzy" universe. The latter one contracts to a state of zero volume and bounces off the singularity. For very special initial conditions the wave function is reflected almost completely and a very small fraction, which will behave pre-classically, enters the domain of positive time. In other words, provided appropriately adjusted initial conditions, the singularity acts as a ``filter", which keeps back all ``fuzzy" impurities and transmits only the purely pre-classical wave function. From the large ratio of the amplitudes of the fuzzy and the smooth wave functions and from the fact that there is no pre-classical wave function at all unless the parameters are very fine tuned, it appears that a (pre-)classical universe is very unlikely to emerge from our model. However, our intuition on probabilities, transition and reflection does not have a reliable framework. To obtain a sound notion of probability and, perhaps, of unitary evolution, one needs a suitable inner product on the space of wave functions. 

The absence of singularities and the possibility to evolve wave functions through a state with zero volume formally solves a problem of classical cosmology and ``standard quantum cosmology", where the scale factor of the universe is considered as continuous. Nevertheless our model does not have a clear interpretation.  In fact it raises much the same kind of questions as standard cosmology. These questions, however, are rephrased in the context of ILQC in a way which might prove productive.

Above all, there is the question of the beginning of the universe. Is a (pre-)classical, expanding universe preceded by a time-reversed, shrinking one? To discuss this question, one has to think about whether in an emphatically non-pre-classical evolution the interpretation of the parameter $n$ as internal time makes sense.  Does the internal time pass from large negative values of $n$ towards zero, or should we ``see" the beginning of the universe at the classical singularity?  Do the positive and negative branches of the wave function describe two possible evolutions of the universe away from the initial zero-volume state? Does the huge ratio between the amplitudes of the non-pre-classical and the pre-classical wave function indicate only a tiny probability for a pre-classical universe to be created?   

The present model poses some interesting questions for ILQC.  One shortcoming (shared with all homogeneous models) is  that the massless, spatially constant scalar field cannot imply the notion of particles, so that it is insufficient for the description of  cosmological particle production.  Another is the issue of stability.  It would be interesting to investigate whether these models would be stable under the inhomogeneous perturbations.  

\subsection*{Acknowledgement}  

The authors thank Martin Bojowald for helpful discussions. F.~H. would like to acknowledge Hamilton College for hospitality and the Czech ministry of education for  support (contract no. 143100006).  S.M. thanks the Perimeter Institute for hospitality and support.  S.M is a Research Corporation Cottrell Scholar.

\end{document}